\documentclass[reprint,twocolumn,superscriptaddress,secnumarabic,amssymb, nobibnotes, aps,pra,superscriptaddress]{revtex4-1}

\usepackage{extarrows}
\usepackage{amsmath}    
\usepackage{graphicx}    
\usepackage{verbatim}    
\usepackage{color}           
\usepackage{subfigure}   
\usepackage{hyperref}    
\usepackage{verbatim}  
\begin{document}

\title{ \textcolor{black}{Charge-transfer effect in hard x-ray 1$s$ and 2$p$ photoemission spectra: \\ LDA+DMFT and cluster-model analysis} }



\author{Mahnaz Ghiasi}
\affiliation{Inorganic Chemistry $\&$ Catalysis, Debye Institute for Nanomaterials Science, Utrecht University, Universiteitsweg 99, 3584 CG, Utrecht, The Netherlands}

\author{Atsushi Hariki}
\affiliation{Institute for Solid State Physics, TU Wien, 1040 Vienna, Austria}
\author{Mathias Winder}
\affiliation{Institute for Solid State Physics, TU Wien, 1040 Vienna, Austria}
\author{Jan Kune\v{s}}
\affiliation{Institute for Solid State Physics, TU Wien, 1040 Vienna, Austria}
\affiliation{Institute of Physics, Czech Academy of Sciences, Na Slovance 2, 182 21 Praha 8, Czechia}

\author{Anna Regoutz}
\affiliation{Imperial College London, Department of Materials, Exhibition Road, London SW7
2AZ, UK}

\author{Tien-Lin Lee}
\affiliation{Diamond Light Source, Didcot, OX11 0DE,UK}

\author{Yongfeng Hu}
\affiliation{Canadian Light Source, Saskatoon, SK S7N 2V3,Canada}

\author{Jean-Pascal Rueff}
\affiliation{Synchrotron SOLEIL, l'Orme des Merisiers, Saint-Aubin, BP 48, F-91192 Gif-sur-Yvette Cedex, France}
\affiliation{Sorbonne Universit\'e, CNRS, Laboratoire de Chimie Physique-Matière et Rayonnement, LCPMR, F-75005 Paris, France}

\author{Frank M. F. de Groot}
\affiliation{Inorganic Chemistry $\&$ Catalysis, Debye Institute for Nanomaterials Science, Utrecht University, Universiteitsweg 99, 3584 CG, Utrecht, The Netherlands}

\date{\today}

\begin{abstract}
We study $1s$ and $2p$ hard x-ray photoemission spectra (XPS) in a series of late transition metal oxides:~Fe$_2$O$_3$ (3$d^{5}$), FeTiO$_3$ (3$d^{6}$), CoO (3$d^{7}$) and NiO (3$d^{8}$). The experimental spectra are analyzed with two theoretical approaches:~MO$_6$ cluster model and local density approximation (LDA) + dynamical mean-field theory (DMFT).
Owing to the absence of the core-valence multiplets and spin-orbit coupling, 1$s$ XPS is found to be a sensitive probe of chemical bonding and nonlocal charge-transfer screening, providing complementary information to  2$p$ XPS.
The 1$s$ XPS spectra are used to assess the accuracy of the $ab$-initio LDA+DMFT approach, developed recently to study the material-specific charge-transfer effects in core-level XPS.
\end{abstract}

\maketitle

\section{\label{sec1}INTRODUCTION}
Transition-metal oxides are an important class of functional materials, showing a variety of fascinating phenomena, such as giant magnetoresistance and spontaneous ordering of spin, charge or orbital degrees of freedom~\cite{imada98,khomskii14}. Thanks to the progress in using the hard x-ray sources in the last decade, core-level x-ray photoemission spectroscopy (XPS) has become a powerful tool to study electronic properties of transition-metal oxides~\cite{Kobayashi09,Taguchi13,Taguchi16book,groot_kotani}. Electronic response to the local charged perturbation (core-hole creation) gives rise to specific features in the $2p$ XPS spectra due to the charge-transfer screening from distant transition metals, traditionally called nonlocal screening~\cite{Veenendaal93,Veenendaal06,Hariki17}, in addition to local screening from  neighboring ligands. The nonlocal screening features are sensitive to various aspects of intersite physics in transition-metal oxides including magnetic and orbital order, and metal-insulator transition~\cite{Horiba04,Veenendaal06,Hariki17,Taguchi08,Eguchi08,Obara10,Kamakura04,Hariki13b,Taguchi05,Chang18}.
\textcolor{black}{The unscreened final states, on the other hand, leave trace of intra-atomic multiplets on the satellite features in the XPS spectra, which are usually well separated in energy from the charge-transfer features~\footnote{\textcolor{black}{The unscreened final states can be described by taking the local core-valence and valence-valence interaction into account, that is encoded in standard impurity-model approaches.}}.}

In 2$p$ XPS, the charge-transfer features are buried in complex spectra reflecting the 2$p$--3$d$ core-valence multiplets and spin-orbit coupling in the 2$p$ shell. These effects are absent in the $1s$ XPS spectra. In practice, little additional effort is required to measure 1$s$ XPS together with valence or other core-level spectra. Despite the large life-time broadening, the absence of the core-valence multiplets and spin-orbit coupling allows 1$s$ XPS to be used:~(1) to identify charge-transfer satellites at higher binding energies, which enables an accurate estimation of material-specific parameters~\cite{Miedema15,Rubio18},~(2) to distinguish local- and nonlocal-screening features. We note that in principle there is a 1$s$--3$d$ exchange interaction but the interaction strength is only a few meV and this effect is not visible in the spectral shape.

\textcolor{black}{The cluster model describing the electronic states of transition-metal ion 
hybridized to discrete states of neighboring ligands
has been widely employed in core-level XPS studies~\cite{groot_kotani,Bocquet92}. The adjustable parameters of the cluster model can be to a large extent eliminated by $ab$-initio low-energy Hamiltonians in the basis of the localized Wannier orbitals~\cite{Haverkort12}. Nevertheless, the cluster model misses the nonlocal charge-transfer screening from the bulk of crystal beyond the ligand atoms. This deficiency can be cured by generalizing the cluster model with discrete ligand states to a quantum impurity model with a continuous spectrum, which represents the host surrounding the excited transition-metal ion. While this description is exact for a noninteracting host, in the cases of our interest, the host is correlated due to electron-electron interaction between localized electrons, even strongly. In that case, dynamical mean-field theory (DMFT)~\cite{metzner1989,georges96} provides an optimized effective description of the host. Following simplified model studies~\cite{Hariki13b,Hariki13a,Hariki15,Hariki16,Haverkort14,Kim04}, $ab$-initio local-density approximation (LDA) approach + DMFT was applied to core-level spectroscopies recently~\cite{Hariki17,Hariki18}, see also Ref.~\cite{Luder17} in a similar direction.
\textcolor{black}{The unified impurity-model description of core-level spectra in a wide range of correlated materials~\cite{Kolorenc92} and in various excitation processes/edges~\cite{Hariki18,Kolorenc18} provided by the LDA+DMFT approach is posted to experimental tests.}}

\textcolor{black}{In this paper, we present a combined experimental and theoretical study of $1s$ and $2p$ XPS spectra in selected late transition-metal oxides:~Fe$_2$O$_3$, FeTiO$_3$, CoO and NiO. By comparison of the cluster-model and LDA+DMFT calculations, we distinguish the local- and nonlocal-screening contributions in the 1$s$ and 2$p$ spectra. We demonstrate that in the 2$p$ spectra, popular choice for the study of 3$d$ transition-metal oxides so far, disentanglement of the local and nonlocal-screening contributions is a complex task due to the core-valence multiplets, 
while asymmetric shape of the main 1$s$ XPS line is found to be a fingerprint of nonlocal screening in the studied compounds. The accuracy of the {\it ab-initio} material-specific parameters is examined by comparing to the present 1$s$ XPS data. The LDA+DMFT approach reproduces both the 1$s$ and 2$p$ spectra consistently.}


\section{\label{sec2}Experimental details}

\textcolor{black}{The Ni 1$s$ XPS of NiO was measured at the HIKE station of the KMC-1 beamline at BESSY.
The photon energy of $\sim$ 8.95~keV was used and the energy resolution is $\sim$ 0.5~eV. According to the Tanuma Powell and Penn algorithm (TPP-2M) formula~\cite{Tanuma94}, the inelastic mean free path (IMFP) for the mentioned photon energy is 13 $\AA$.
The NiO thin film was grown on a Ag substrate~(001) and capped by 3~nm of MgO to avoid the charging effects~\cite{Calandra12}}.

\textcolor{black}{The Co 1$s$ and 2$p$ XPS spectra of the single crystal CoO were measured at SXRMB beamline of the Canadian Light Source (CLS). The photon energy of 9 (3) keV for 1$s$ (2$p$) were used with the energy resolution of 0.9 (0.3) eV. The IFMP (obtained with TPP-2M) is estimated as  23 (35) $\AA$ for 1$s$ (2$p$) measurements.}

\textcolor{black}{The Fe 1$s$ and 2$p$ XPS spectra of $\alpha$-Fe$_2$O$_3$ single crystal were measured at ID16 beamline of ESRF. The photon energy of 10~keV and 7.7~keV were used for the 1$s$ and 2$p$ measurement, with the IMFP (obtained with TPP-2M) of 50 and 105 $\AA$, respectively. The energy resolution at 10~keV is 0.48~eV and at 7.7~keV is 0.42~eV.}

\textcolor{black}{The Fe 1$s$ XPS and 2$p$ XPS spectra of FeTiO$_3$ (Fe$^{2+}$) powders (99.8\%) were measured at I09 beamline of Diamond light source. The photon energy of 12~keV and 6~keV were used for the 1$s$ and 2$p$ measurements, respectively. The energy resolution at 12~keV is 0.3~eV and at 6~keV is 0.25~eV. The IFMP (obtained with TPP-2M) is estimated as 93 and 100 $\AA$ for the 1$s$ and 2$p$ photon energy, respectively.}

\textcolor{black}{All experimental data were collected at room temperature.}

\textcolor{black}{We present fitting analysis for the experimental 1$s$ spectra. The main line is fitted using two Pseudo-Voigt functions composed of 20\% Lorentzian and 80\% of Gaussian. Considering the present experimental energy resolution and typical life-time of the 1$s$ core hole, the width (HWHM) of the Lorentzian $\Gamma_{\rm L}$ and Gaussian $\Gamma_{\rm G}$ is set to ($\Gamma_{\rm L}$, $\Gamma_{\rm G}$)=(0.08, 1.02), (0.28, 1.12), (0.29, 0.91), and (0.20, 0.80) in the unit of eV for NiO, CoO, Fe$_2$O$_3$, and FeTiO$_3$, respectively.
For the satellites, the widths of the Pseudo-Voigt functions are relaxed to reproduce the experimental data. The background in the experimental data is subtracted prior to the fitting analysis using the Shirley method~\cite{Shirley72}.}

\section{\label{sec3}COMPUTATIONAL METHOD}

\subsection{\label{sec3a}Theoretical Methods}

The $1s$ and $2p$ XPS spectra are analyzed by the cluster model and the LDA+DMFT approach. \textcolor{black}{Both approaches build on the same lattice model constructed from LDA calculations but use different approximations for describing the valence states surrounding the x-ray excited transition-metal site.}
Both approaches start with a standard LDA calculation using the Wien2K package~\cite{wien2k} for the experimental crystal structure. Then the tight-binding representation of the transition metal 3$d$ and O 2$p$ bands is constructed using the Wannier90 package~\cite{wannier90} and Wien2wannier interface~\cite{wien2wannier}. The $d$--$p$ model is augmented with the electron--electron interaction within the transition metal 3$d$ shell, leading to the lattice Hamiltonian
\begin{eqnarray}
 H&=&\sum_{\textit{\textbf{k}}} 
  \begin{pmatrix}
      \textit{\textbf{d}}^{\dag}_\textit{\textbf{k}} & \textit{\textbf{p}}^{\dag}_\textit{\textbf{k}}
 \end{pmatrix}
 \begin{pmatrix}
     h^{dd}_{\textit{\textbf{k}}}-\mu_{\rm dc} & h^{dp}_{\textit{\textbf{k}}} \\
     h^{pd}_{\textit{\textbf{k}}} & h^{pp}_{\textit{\textbf{k}}}
 \end{pmatrix}
   \begin{pmatrix}
      \textit{\textbf{d}}_\textit{\textbf{k}} \\
      \textit{\textbf{p}}_\textit{\textbf{k}}
 \end{pmatrix} \notag \\
 &+&\sum_{i}\sum_{\kappa\lambda\mu\nu}U_{\kappa\lambda\mu\nu}
    d^\dag_{\kappa i}d^\dag_{\lambda i}
    d^{\phantom{\dag}}_{\nu i}d^{\phantom{\dag}}_{\mu i}.
    \label{ham_int}
\end{eqnarray}
Here, $\textit{\textbf{d}}^{}_\textit{\textbf{k}}$ ($\textit{\textbf{p}}^{}_\textit{\textbf{k}}$)
is an operator-valued vector whose elements
are Fourier transforms of $d_{\gamma i}$  ($p_{\gamma i}$), that annihilates the
transition metal 3$d$ (O 2$p$) electron in the \textcolor{black}{flavor $\gamma$ (combined orbital and spin indices)} in the $i^{th}$ unit cell.
\textcolor{black}{The local Coulomb interaction, acting on the transition metal $d$ shell, is parameterized by Slater integrals $F_0$, $F_2$ and $F_4$. We fix the ratio $F_4/F_2= 0.625$, that enables to determine Coulomb vertices $U_{\kappa\lambda\mu\nu}$ using Habbard $U=F_0$ and Hund's $J=(F_2+F_4)/14$ parameters~\cite{Pavarini1}.}
The double-counting term $\mu_{\rm dc}$, which corrects for the $d$--$d$ interaction
presented in the LDA step, renormalizes the $p$--$d$ splitting, thus relates to the charge-transfer energy $\Delta$~\cite{Hariki17,Hariki18,Karolak10}.
We introduce the relation of $\Delta$ and $\mu_{\rm dc}$ as, $\Delta = \varepsilon_d -\mu_{dc}+n\times U_{dd} - \varepsilon_p$, where $\varepsilon_d$ ($\varepsilon_p$) is the average energy of transition metal 3$d$ (O 2$p$) states and  $U_{dd}$ is the value of the (configuration-averaged) Coulomb interaction.
The calculation temperature is set to $T=300$~K in the present study.
We note that, in NiO, CoO and Fe$_2$O$_3$, the unit cell is enlarged to simulate the antiferromagnetic ordering observed experimentally below 300~K~\footnote{In the LDA+DMFT calculation, the self-energy depends on the spin in the AF phase.}. In FeTiO$_3$, the unit cell contains Ti and Fe atoms.

\textcolor{black}{In the cluster-model approach, following the conventional definition, we discard all states in Eq.~(1) except for those on the x-ray excited transition-metal and its nearest-neighboring ligands. The waverfunction of the initial state as well as the final states is represented in the truncated Fock space. Thus by construction nonlocal charge transfer beyond neighboring ligands cannot be described by the cluster model. In the LDA+DMFT approach, on the other hand, we first solve the lattice model of Eq.~(1) within DMFT~\cite{georges96,kotliar06}.
The DMFT renders local self-energy to one-particle LDA bands, that yields many-body description of valence excitations.
The infinite lattice model is mapped onto the Anderson impurity model coupled to the optimized non-interacting host that represents valence states around the excited metal.
The Hamiltonian of the cluster model is given by
\begin{eqnarray}
  H_{\rm clu}=H_{\rm TM}
  +\sum_{\gamma}\varepsilon^p_{\gamma} 
  \tilde{p}_{\gamma}^\dagger \tilde{p}_{\gamma} 
  +\sum_{\gamma} t_{\gamma}(d_{\gamma 0}^\dagger \tilde{p}_{\gamma} + h.c), \notag
\end{eqnarray}
where $t_{\gamma}$ represents the hybridization intensity between the metal $d$ state and molecular orbital $\tilde{p}_{\gamma}$   (with energy $\varepsilon^p_{\gamma}$) composed of 2$p$ states on nearest neighboring ligands. The Hamiltonian of the Anderson impurity model in the LDA+DMFT approach is given by
\begin{eqnarray}
  H_{\rm and}=H_{\rm TM}+
  \sum_{\alpha\gamma} \epsilon_{\alpha\gamma} v_{\alpha\gamma}^\dagger v_{\alpha\gamma} +  \sum_{\alpha\gamma} V_{\alpha\gamma}(d_{\gamma 0}^\dagger v_{\alpha\gamma} + h.c), \notag
\end{eqnarray}
where ${v_{\alpha\gamma}^{\, \dagger}}$ (${v_{\alpha\gamma}}$) is the creation (annihilation) operator for the auxiliary state $\alpha$ with energy $\epsilon_{\alpha\gamma}$ in the DMFT hybridization function.
The third term discriminates the cluster model and LDA+DMFT approach in the description of the hybridization between the excited metal and the rest of the crystal.
In the LDA+DMFT approach, the effect of the long-distant atoms is represented by hopping amplitude $V_{\alpha\gamma}$.
The two approaches have the same form of the local terms acting on the excited metal site
\begin{eqnarray}
H_{\rm TM}&=&\sum_{\gamma\sigma}\tilde{\varepsilon}^d_{\gamma} d_{\gamma 0}^{\, \dagger} d_{\gamma 0}
+\sum_{\kappa\lambda\mu\nu}U_{\kappa\lambda\mu\nu}
    d^\dag_{\kappa 0}d^\dag_{\lambda 0}
    d^{\phantom{\dag}}_{\nu 0}d^{\phantom{\dag}}_{\mu 0}
    \notag \\
 &-&U_{dc} \sum_{\gamma \eta} 
  d_{\gamma0}^{\, \dagger} d_{\gamma0} (1-c_{\eta0}^{\, \dagger} c_{\eta0}) 
 +H_{\rm multiplet}. \notag
\end{eqnarray}
Here, $c_{\eta0}^{\, \dagger}$ is the creation operator for core electron (1$s$ or 2$p$). The on-site energies of transition-metal $d$ states $\tilde{\varepsilon}^d_{\gamma}=\varepsilon^{d}_{\gamma}-\mu_{\rm dc}$ are the ones
of the wannier states $\varepsilon^{d}_{\gamma}$ shifted by the double-counting correction $\mu_{\rm dc}$. 
The isotropic part of the core-valence interaction $U_{dc}$ is shown explicitly,
while other terms containing higher multipole contributions and the spin-orbit interaction are included in $H_{\rm multiplet}$, see Sec.~IIIB. Since common basis functions ($d$~wannier functions) are adopted in the cluster model and LDA+DMFT approach, we use the same interaction parameters in the two, actual values are given in Sec.~III\ref{sec3b}. }



\textcolor{black}{In the present LDA+DMFT calculations,}
the strong coupling continuous-time quantum Monte Carlo method~\cite{werner06,gull11,boehnke11,hafermann12} with density-density approximation to \textcolor{black}{the Coulomb vertices in Eq.~(1)} was used to compute the self-energy of the auxiliary Anderson impurity model. Upon reaching the LDA+DMFT self-consistency, the self-energy is analytically continued with the maximum entropy method~\cite{wang09,jarrell96} in order to obtain the hybridization function $V(\varepsilon)$ in the real-frequency domain. Then,~\textcolor{black}{for computing core-level spectra,} we augment the Anderson impurity model with the core orbitals ($1s$ or $2p$).

The core-level XPS spectrum is calculated with the truncated configuration interaction method~\cite{Hariki17}. The XPS spectral function for the binding energy $E_{\rm B}$ is given by~\cite{groot_kotani,Hariki17}
\begin{eqnarray}
F^{(n)}_{\rm XPS}(E_{\rm B})=-\frac{1}{\pi}\sum_{n} \langle n | T_{D}^{\dag} \frac{1}{E_{\rm B}+E_n-H_{\rm model}} T_D | n \rangle \notag,
\end{eqnarray}
where $E_n$ is the eigen energy of the $n$-th excited states $|n\rangle$ in the initial state.
The temperature effect is taken into account with the Boltzmann factor at $T$~=~300~K.
The operator \textcolor{black}{$T_D$} creates a 1$s$ or 2$p$ core hole at the impurity transition-metal site. Here $H_{\rm model}$ is the model~Hamiltonian, i.e.~the~Hamiltonian of the cluster model 
\textcolor{black}{$H_{\rm clu}$}
or the Anderson impurity model
\textcolor{black}{$H_{\rm and}$}
with the DMFT hybridization function.
\textcolor{black}{In both models, full Coulomb vertices for the valence-valence and core-valence interaction are included in the XPS calculations explicitly.}
The computational details in the LDA+DMFT method can be found in Refs.~\cite{Hariki17,Hariki18}.
\textcolor{black}{In the cluster-model approach, we employ the CTM4XAS program~\cite{Frank10}, which implements a standard three configuration scheme.}


\begin{table}[t]
	\begin{ruledtabular}
		\begin{tabular}{l c c c c}
			& Fe$_2$O$_3$ & FeTiO$_3$ & CoO & NiO \\
			& Fe$^{3+}$ ($d^5$) & Fe$^{2+}$ ($d^6$) & Co$^{2+}$ ($d^7$) & Ni$^{2+}$ ($d^8$) \\
			\hline

			10$Dq$  & 0.5 & 0.25 & 0.25 & 0.45  \\
			$\Delta$  & 3.7 & 3.5 & 4.1 & 4.4  \\
			$U_{sd}$ ($U_{pd}$)  & 8.4 & 8.0 & 8.6 & 7.8 \\
			$U_{dd}$ & 6.4 & 6.4 & 6.8 & 6.5 \\
			V$_{T_{2g}}$ & 1.3 & 1.3 & 1.2 & 1.2 \\
			V$_{E_g}$ & 2.5 & 2.1 & 2.0 & 2.1 \\
			$\Gamma_{1s}$ & 0.7 & 1.1 & 0.9 & 0.9 \\
			$\Gamma_{2p}$ & 0.5 & 0.5 & 0.4 & 0.4 \\
		\end{tabular}
	\end{ruledtabular}
		\caption{ Summary of the parameter values in the present study:~crystal-field splitting between $E_g$ and $T_{2g}$ states ($10Dq$), charge-transfer energy ($\Delta$), core-hole potential for 1$s$ ($U_{sd}$) and 2$p$ ($U_{pd}$) XPS, 3$d$ Coulomb interaction ($U_{dd}$), hopping amplitude of the nearest-neighboring ligand states and transition metal $E_g$ ($V_{E_g}$) and $T_{2g}$ ($V_{T_{2g}}$) state.
		\textcolor{black}{$\Gamma_{1s}$ and $\Gamma_{2p}$ are the Gaussian broadening (half width at half maximum) included in theoretical 1$s$ and 2$p$ spectra, respectively.} In Fe$_2$O$_3$ and FeTiO$_3$, the triply-degenerate $t_{2g}$ states split into double-degenerate $e_{g\pi}$ states and single $a_{1g}$ state. The values of the $t_{2g}$ state in the table are obtained by averaging over the ones of the $e_{g\pi}$ and $a_{1g}$ states. In actual calculation, the splitting of the $e_{g\pi}$ and $a_{1g}$ states is taken into account explicitly. All values are in eV.
		}\label{tab3}
\end{table}

\subsection{\label{sec3b}Computational Parameters}

The following computational parameters are used both in the cluster model and LDA+DMFT method:~the Coulomb interaction of the 3$d$ electrons $U_{dd}$;~the core-valence Coulomb interactions $U_{sd}$ ($U_{pd}$) in the $1s$ ($2p$) XPS;~Slater integrals representing the multipole part of the Coulomb interaction;~one-particle hopping parameters;~crystal-field splitting;~and charge-transfer energy $\Delta$. 
The $U_{dd}$ value is fixed by consulting with DFT-based estimations or previous XPS studies, as given in~Table~\ref{tab3}.
The core-hole potential $U_{sd}$ ($U_{pd}$) is fixed by fitting the experimental core-level spectra.
\textcolor{black}{Based on experimental observations in Sec.~IV, we use the same value for $U_{sd}$ and $U_{pd}$.}
\textcolor{black}{The multipole part of the core-valence interaction is determined by the Slater integrals $F_k$ and $G_k$, defined as~\cite{groot_kotani}}
\begin{eqnarray}
 F_k&=&\int_{0}^{\infty} \int_{0}^{\infty} R^2_{\eta}(r_1)R^2_{3d}(r_2) \frac{r^k_{<}}{r^{k+1}_{>}}r^2_{1}r^2_{2}\ dr_1 dr_2 \notag \\
 G_k&=&\int_{0}^{\infty} \int_{0}^{\infty}
 R_{\eta}(r_1) R_{3d}(r_1)R_{\eta}(r_2) R_{3d}(r_2) \notag \\
 &\times&   \frac{r^k_{<}}{r^{k+1}_{>}}r^2_{1}r^2_{2} \ dr_1 dr_2, \notag
\end{eqnarray}
\textcolor{black}{where $\eta$ denotes $1s$ and $2p$ orbit for 1$s$-3$d$ and 2$p$-3$d$ core-valence interactions, respectively.
Here $r_{>}$ ($r_{<}$) means the larger (smaller) one of $r_1$ and $r_2$, and $R_{nl}(r)$ is corresponding radial function.}
The spin-orbit coupling in the 2$p$ and 3$d$ shell, and the Slater integrals $F_k$, $G_k$ are calculated with an atomic Hartree-Fock code. The $F_k$ and $G_k$ values are scaled down to 80\% of the Hartree-Fock values to simulate the effect of intra-atomic configuration interaction from higher basis configurations, which is a successful empirical treatment~\cite{Matsubara05,Sugar72,Tanaka92,Groot90,Cowan18}.
The one-particle hopping and the crystal-field parameters are obtained from the LDA bands. 
The double-counting correction $\mu_{\rm dc}$ in LDA+DMFT is treated as an adjustable parameter and fixed by comparison to the valence photoemission spectra~\cite{Hariki17}.
For FeTiO$_3$, $\mu_{dc}$ is determined to reproduce the experimental gap ($\sim 2.5$~eV)~\cite{Ginley77} since valence photoemission data are not reported so far.
The computed XPS intensities are broadened using Gaussian function to simulate the instrumental resolution and the finite core-hole life time.

 \subsection{\label{sec3c}Hybridization Function}
 
 The hybridization function $V(\varepsilon)$ describes electron hopping between the 3$d$ orbitals on the x-ray excited transition-metal site and the rest of the crystal. 
 In the cluster model, which is a special case of the Anderson impurity model, the hybridization function has a form of Dirac delta function and usually is not introduced
 explicitly. The LDA+DMFT hybridization function is a general function of energy $\varepsilon$, which is obtained for \textcolor{black}{flavor $\gamma$}~\footnote{In the present study, we neglect the small off-diagonal elements of $V_{\gamma}^2(\varepsilon)$. },
\begin{equation}
   V_{\gamma}^2(\varepsilon)=-\frac{1}{\pi}{\rm Im}
        \langle d_{\gamma}|\bigl (\varepsilon-h^{0}-
        \Sigma(\varepsilon)-G^{-1}(\varepsilon)\bigr )^{-1}
|d_{\gamma}\rangle,
\end{equation}
where $\Sigma(\varepsilon)$ and $h^0$ are 
the local self-energy and the one-body part of the on-site Hamiltonian, respectively~\cite{georges96,kotliar06,Hariki17}.
The local Green's function $G(\varepsilon)$ is computed by averaging the lattice Green's function over the Brillouin zone. While the Anderson impurity model provides only an approximate description of local dynamics in interacting lattice system, it becomes exact
for non-interacting host. In this case, the hybridization function $V_{\gamma}(\varepsilon)$ can be decomposed to the distance-shell contributions,
\begin{equation}
   V_{\gamma}^2(\varepsilon)=-\frac{1}{\pi}{\rm Im}\sum_{\mu,\nu}
      \langle \gamma| \operatorname{V}^{\dag} |\mu\rangle\langle\mu|
      \frac{1}{\varepsilon-h_{\rm b}} 
       |\nu\rangle\langle\nu|
      \operatorname{V}
|\gamma\rangle.
\end{equation}
\textcolor{black}{Here operator $\operatorname{V}$ describes the hopping between the transition metal 3$d$ orbitals on the impurity site and the rest of the crystal.}
\textcolor{black}{The operator $h_{\rm b}$ is the host Hamiltonian which contains all terms in Eq.~(1) except the ones including the x-ray excited transition-metal site. The indices $\mu$ and $\nu$ runs over the lattice sites
and thus the summation can be truncated at the desired distance.}
\textcolor{black}{For example, the hybridization function of the cluster model (with $H_{\rm clu}$ in Sec.~IIIA) is given as,}
\begin{equation}
   V_{\gamma}^2(\varepsilon)= |t_{\gamma}|^2 \times \delta(\varepsilon - \varepsilon^p_{\gamma}) \notag.
\end{equation}
\textcolor{black}{We use Eq.~(3) to compute $V_{\gamma}(\varepsilon)$ of the cluster model as well as the (non-interacting) finite-size clusters to demonstrate hopping-distance dependence of $V_{\gamma}(\varepsilon)$ structure.
Analysis of the contributions of the various distance-shells will be discussed later in Fig.~\ref{fig_demo}.
Note that similar expression for $V_{\gamma}^2(\varepsilon)$ holds also for an interacting host (with an additional self-energy in the denominator). 
However, decomposition to different hopping cut-offs is not strictly defined since all hoppings are implicitly present in the self-energy.}

\section{\label{sec4} RESULTS}

\subsection{\label{sec4_1} NiO}

\begin{figure}[t]
	\graphicspath{{Main/}}
	\includegraphics[width=1.00\columnwidth]{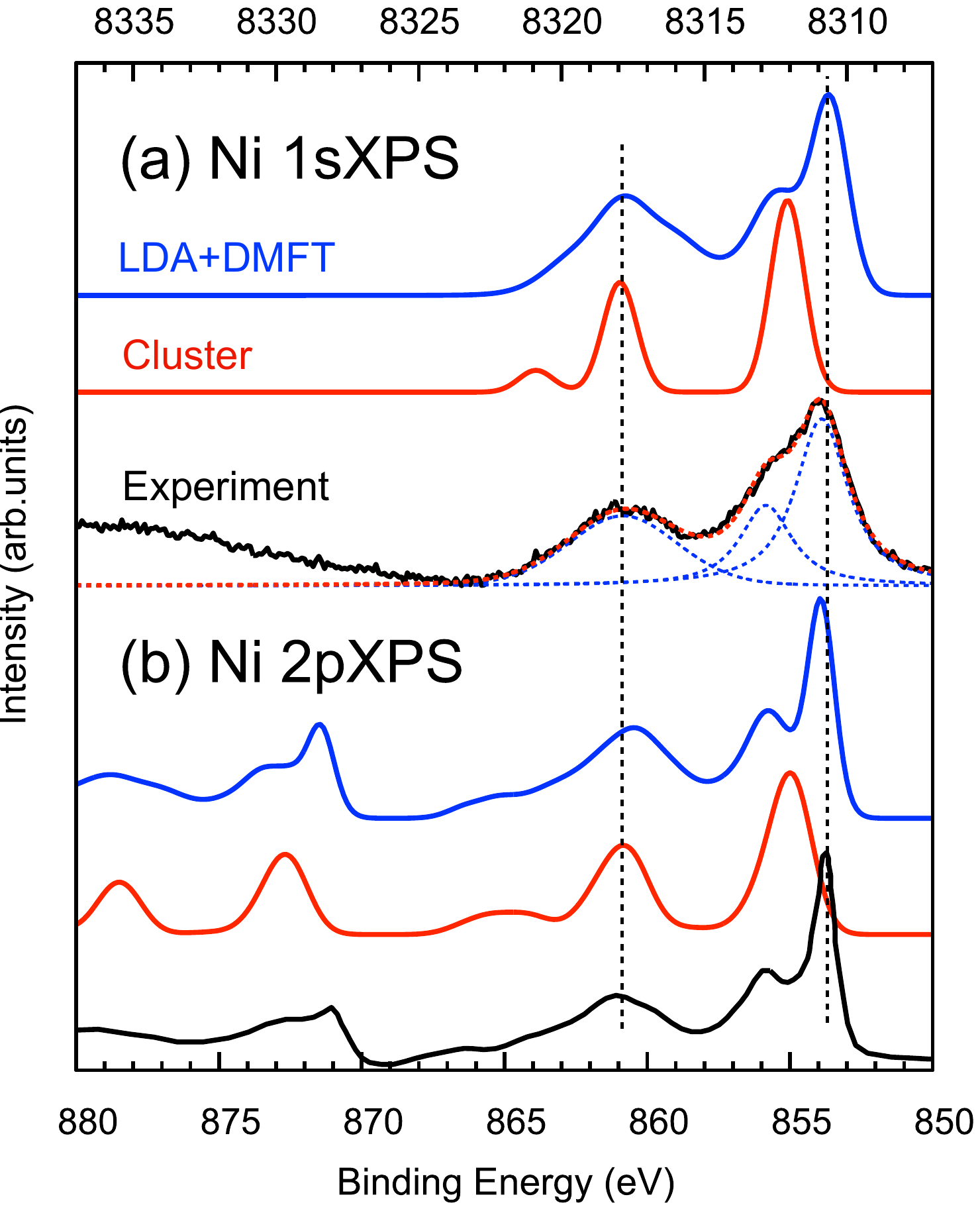}
	\caption{\textcolor{black}{(Color online) Experimental Ni 1$s$ and 2$p$  XPS spectra of NiO (black) are compared with LDA+DMFT in the antiferromagnetic phase (black, solid) and cluster-model (red, solid) calculations. The Gaussian spectral broadening of 0.7 (0.5) eV (HWHM) is taken into account in the calculated 1$s$ (2$p$) spectra. The experimental Ni 2$p$ XPS data is taken from Ref.~\cite{Taguchi08}. The fitting result (red, dashed) using Voigt functions (black, dashed) for the 1$s$ data is shown together.} 
	}\label{fig_ni}
\end{figure}

\textcolor{black}{We start with NiO, a prototype Mott insulator  with a large charge gap ($\sim$~4~eV). Figure~\ref{fig_ni} shows Ni 1$s$ and 2$p$ XPS spectra of NiO, together with the calculated spectra by the cluster model and LDA+DMFT. The 1$s$ main line around 8311~eV corresponds to $\underline{c}d^9\underline{v}$ configuration, where $\underline{c}$ and $\underline{v}$ denote a 1$s$ core-hole and a hole in the valence band, respectively. In addition, the charge-transfer satellite with a mixed character of $\underline{c}d^8$ and $\underline{c}d^{10}\underline{v}^2$ configurations is clearly observable around 8318~eV. The Ni 1$s$ XPS spectrum is free from spin-orbit coupling in the core level, while the 2$p$ spectrum is composed of 2$p_{3/2}$ (868$\sim$853~eV) and 2$p_{1/2}$ (885$\sim$870~eV) components. Since the spin-orbit coupling in the Ni 2$p$ core level is large ($\sim$11~eV), the Ni 2$p_{3/2}$ and 2$p_{1/2}$ components have no overlap. The 2$p_{3/2}$ ($2p_{1/2}$) main line locates around 854~eV (873~eV) and the 2$p_{3/2}$ ($2p_{1/2}$) charge-transfer satellite is observed around 861~eV (878~eV).
The splittings of the main line and the charge-transfer satellite in the Ni 1$s$ and 2$p$ spectra are almost identical to each other ($\sim 6$~eV), indicating the values of the core-hole potential $U_{sd}$ and $U_{pd}$ are comparable. Indeed, the LDA+DMFT calculation with $U_{sd}=U_{pd}=7.8$~eV well reproduces the splitting of the main and charge-transfer satellite in both the 1$s$ and 2$p$ XPS spectra.
A double-peak feature is observed in the 2$p_{3/2}$ main line. The lower (higher) binding-energy $E_{B}$ side of the double peaks is due to the nonlocal (local) screening in the final states~\cite{Taguchi08,Hariki13b,Hariki17}. The LDA+DMFT result qualitatively reproduces the Ni 2$p$ XPS data including the double-peak lacked in the cluster-model result. We find that a double peak is discernible in the 1$s$ main line despite the larger core-hole broadening. The similarity to the 2$p_{3/2}$ main line and its presence/absence in the LDA+DMFT/cluster-model spectra suggests its nonlocal screening origin. In addition, the charge-transfer satellite shows a noticeable difference in the the cluster and LDA+DMFT results, indicating the nonlocal screening affects not only the main line but also the satellite with a higher binding energy. This is because the charge-transfer satellite has a contribution of the $cd^{10}\underline{v}^2$ configuration, the so-called over-screened states, in which the nonlocal screening takes part. }

\subsection{\label{sec4_2} CoO}

\begin{figure}[t]
	\graphicspath{{Main/}}
	\includegraphics[width=1.00\columnwidth]{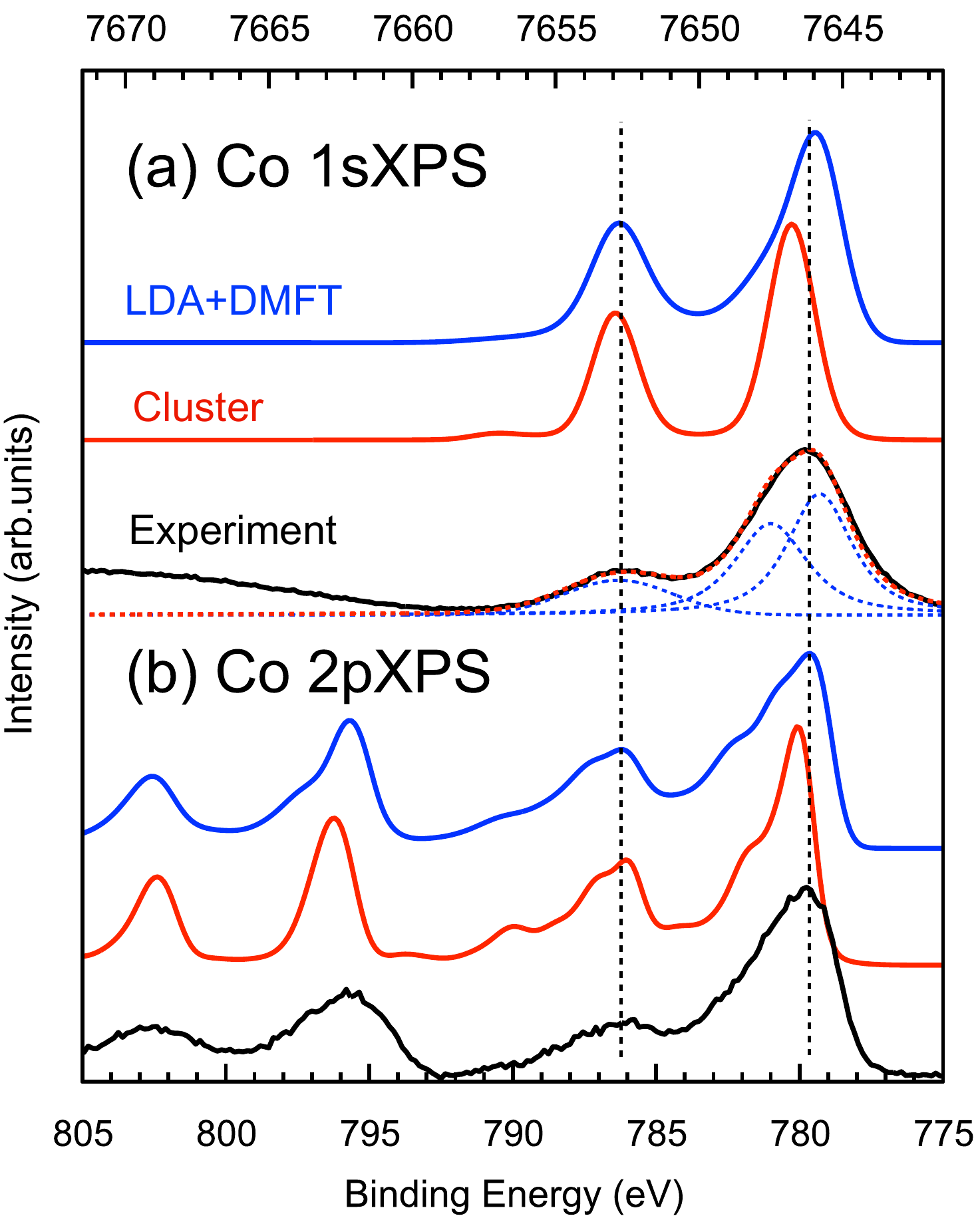}
	\caption{\textcolor{black}{(Color online) Experimental Co 1$s$ and 2$p$  XPS spectra of CoO (black) are compared with LDA+DMFT (black, solid) and cluster-model (red, solid) calculations. The Gaussian spectral broadening of 1.1 (0.5) eV (HWHM) is taken into account in the calculated 1$s$ (2$p$) spectra. The fitting result (red, dashed) using Voigt functions (black, dashed) for the 1$s$ data is shown together.}
    }\label{fig_co}
\end{figure}

\textcolor{black}{CoO is a typical transition-metal oxide with a large charge gap $\sim 3.6$~eV.
Figure~\ref{fig_co} shows Co 1$s$ and 2$p$  XPS spectra of CoO, together with the calculated spectra by the cluster model and LDA+DMFT. 
The Co 1$s$ XPS, see Fig.~\ref{fig_co}a, shows the main line ($\sim$7647~eV) corresponding to $\underline{c}d^8\underline{v}$ configuration and the charge-transfer satellite ($\sim 7653$~eV). 
In the 2$p$ spectra, 2$p_{3/2}$ (2$p_{1/2}$) component is located in 794-778~eV (807-795~eV), in which 
the main line and satellite are observed around 780~eV (796~eV) and 787~eV (803~eV), respectively.
The splitting of the main line and the charge-transfer satellite in the 1$s$ and 2$p$ spectra is almost identical ($\sim$6~eV). 
The 1$s$ main line has an asymmetric shape with a shoulder on the higher $E_B$ side. Because of the absence of the core-valence multiplets in Co 1$s$ spectra, the Co 1$s$ main line is expected to be a single peak. Indeed, the cluster-model calculation yields a symmetric main line. On the other hand, the LDA+DMFT spectrum contains asymmetric main line, suggesting the nonlocal screening is the origin of the asymmetry of the main line.
Then, in Co 2$p$ spectra, we find the 2$p_{3/2}$ main line is rather broad, in a clear contrast to the Ni 2$p_{3/2}$ main line of NiO. The LDA+DMFT result well reproduces the broad shape of the main line as well as of the charge-transfer satellite, compared to the cluster-model result. The difference between the LDA+DMFT and cluster-model results suggests that the nonlocal screening from Co 3$d$ bands plays a role in the formation of the broad asymmetric main line~\cite{Hariki17}. However, the Co 2$p_{3/2}$ main line in the cluster-model result has inner features due to rich 2$p$-3$d$ core-valence multiplets in the $\underline{c}d^8\underline{v}^1$ configuration.
Thus a theoretical simulation is required to disentangle the local screening and nonlocal screening contributions in the Co 2$p$ spectra~\cite{Hariki17}.}

\subsection{\label{sec4_3} Fe$_2$O$_3$}

\begin{figure}[t]
	\graphicspath{{Main/}}
	\includegraphics[width=1.00\columnwidth]{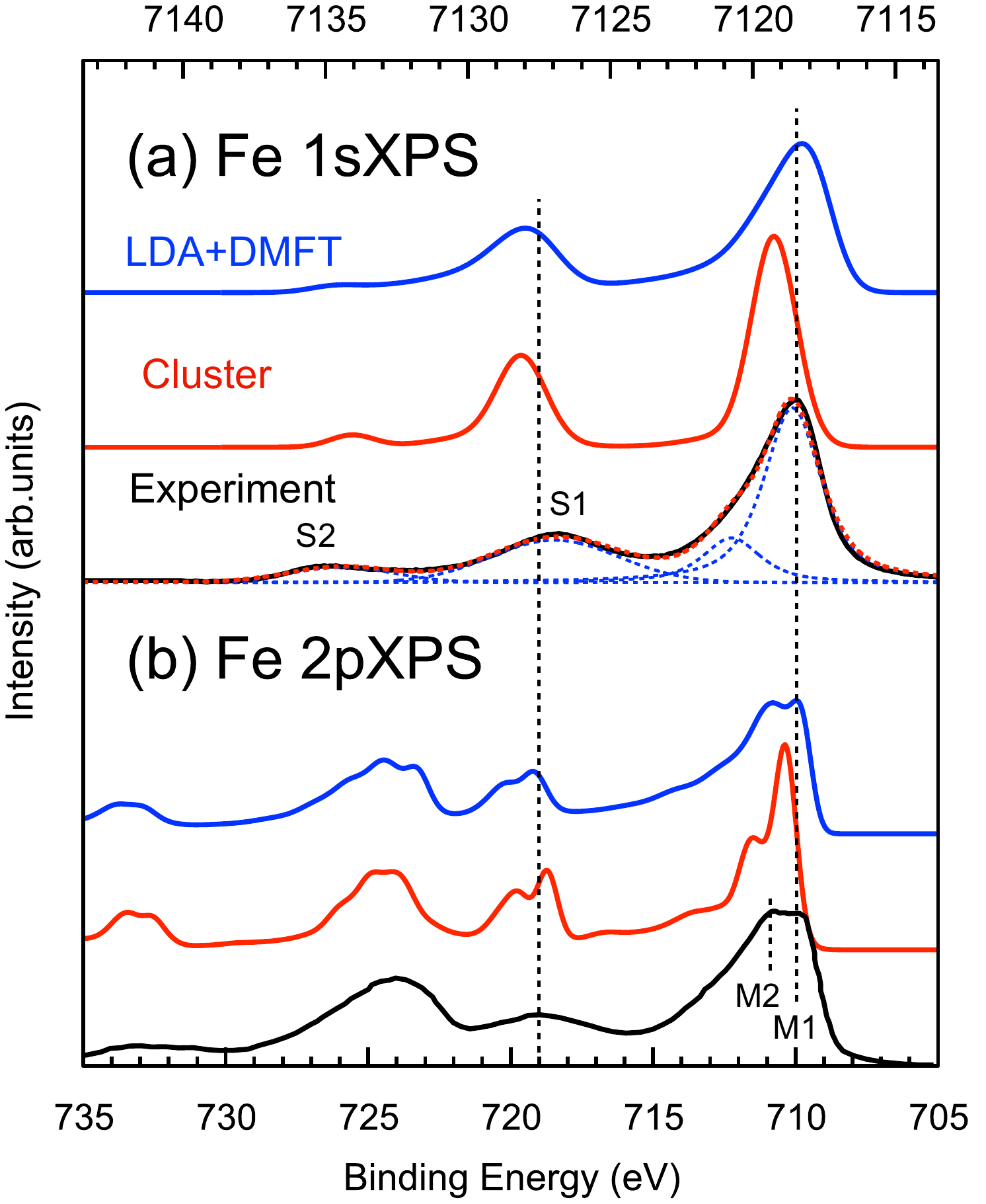}
	\caption{\textcolor{black}{(Color online) Experimental Fe 1$s$ and 2$p$  XPS spectra of Fe$_2$O$_3$ (black) are compared with LDA+DMFT (black, solid) and cluster-model (red, solid) calculations. The spectral broadening using a Gaussian of 0.9 (0.4) eV width (HWHM) is taken into account in the calculated 1$s$ (2$p$) spectra. The experimental data of Fe 2$p$  XPS is taken from Ref.~\cite{Miedema15}. In the 1$s$ spectra, the first and second satellites are labeled as S1 and S2 in Fig.~\ref{fig_fe}a, respectively.
	The fitting result (red, dashed) using Voigt functions (black, dashed) for the 1$s$ data is shown together.}
	}\label{fig_fe}
\end{figure}

$\alpha$-Fe$_2$O$_3$ is an insulating Fe$^{3+}$ oxide with corundum structure and antiferromagnetic ordering below $T_{\rm N}\sim$ 950~K. 
Figure~\ref{fig_fe} shows Fe 1$s$ and 2$p$  XPS spectra, together with the calculated spectra by LDA+DMFT and cluster model.
The Fe 1$s$ spectrum shows three peaks:~main line ($\sim$ 7118.5~eV), the first satellite (S1~:~$\sim$ 7128~eV) and the second satellite (S2~:~$\sim$ 7135~eV). The energy splittings of the main line and satellites are rather large ($\sim$~9.5~eV for S1 and $\sim$~17~eV for S2) compared to those in NiO and CoO ($\sim$~6~eV).  
The large splitting in Fe$_2$O$_3$ can be explained by the value of the effective hybridization $V_{\rm eff}$~\cite{groot_kotani,Uozumi96,Uozumi97},
\begin{equation}
  V_{\rm eff}=\sqrt{ ( 4- N_{\rm E_g} )\times V^2_{\rm E_g}+ ( 6- N_{\rm T_{2g} } )\times V^2_{\rm T_{2g} }},  \notag
\end{equation}
where $N_{\rm E_g}$ ($N_{\rm T_{2g}}$) and $V_{\rm E_g}$ ($V_{\rm T_{2g}}$) are the occupation of the $E_g$ ($T_{2g}$) states and the (bare) hybridization intensity between ligand and the $E_g$ ($T_{2g}$) orbitals~\footnote{For simplicity, octahedral symmetry is assumed and the effect of long-distance hopping (nonlocal screening) is not taken into account in the qualitative discussion.}.
The $V_{\rm eff}$ values in NiO, CoO and Fe$_2$O$_3$, computed for high-spin ground state in formal valence, are 2.97, 3.24, and 4.19~eV, respectively.   
Thus the different configurations ($d^n$, $d^{n+1}\underline{L}$, $d^{n+2}\underline{L}^2$, here $\underline{L}$ denotes a hole in nearest neighboring ligands) are split more in Fe$_2$O$_3$ as compared to NiO and CoO, yielding the large separations of the main line and satellites in Fe 1$s$ XPS. 
Thus in Fe and earlier transition-metal oxides~\cite{Uozumi96,Uozumi97,Bocquet96}, the hybridization strength between transition metal 3$d$ and surrounding atoms can be estimated accurately by the satellite positions since the large $V_{\rm eff}$ magnifies its bare value.
Both the LDA+DMFT and the cluster-model calculations reproduce the positions of the satellites reasonably well, reflecting the accuracy of hopping parameters obtained from the LDA calculation.
However, in the Fe 2$p$ spectrum, Fig.~\ref{fig_fe}b, the second satellite S2 is not visible due to its overlap with the main line of the Fe 2$p_{1/2}$ component ($\sim$ 725~eV).
Thus, thanks to the absence of spin-orbit coupling, 1$s$ XPS complements the 2$p$ XPS information about bonding.

The Fe 2$p_{3/2}$ main line shows a double-peak shape, marked as M1 and M2 in the figure, that is well reproduced in the LDA+DMFT result. 
The difference in the LDA+DMFT and cluster-model results is attributed to the contribution of the nonlocal screening, the M2 intensity is enhanced relative to the M1 one. 
However, as seen in the cluster-model spectrum, Fig.~\ref{fig_fe}b, the main line has a rich fine structure also due to the core-valence Coulomb multiplets, 
which makes determination of the nonlocal screening contribution a difficult task.
On the contrary, the asymmetry of the Fe 1$s$ XPS main line, observed in the experiment, Fig~\ref{fig_fe}a, is solely due to nonlocal screening.
As in CoO, the 1$s$ main line of the cluster model consists of a single peak due to the absence of the core-valence multiplets, while that of the LDA+DMFT shows a clear asymmetry. Thus the shape of the 1$s$ XPS main line provides an unambiguous signature of the nonlocal screening, while it is hidden in the complex structure of 2$p$ XPS.

\subsection{\label{sec4_4} FeTiO$_3$}

\begin{figure}[t]
	\graphicspath{{Main/}}
	\includegraphics[width=1.00\columnwidth]{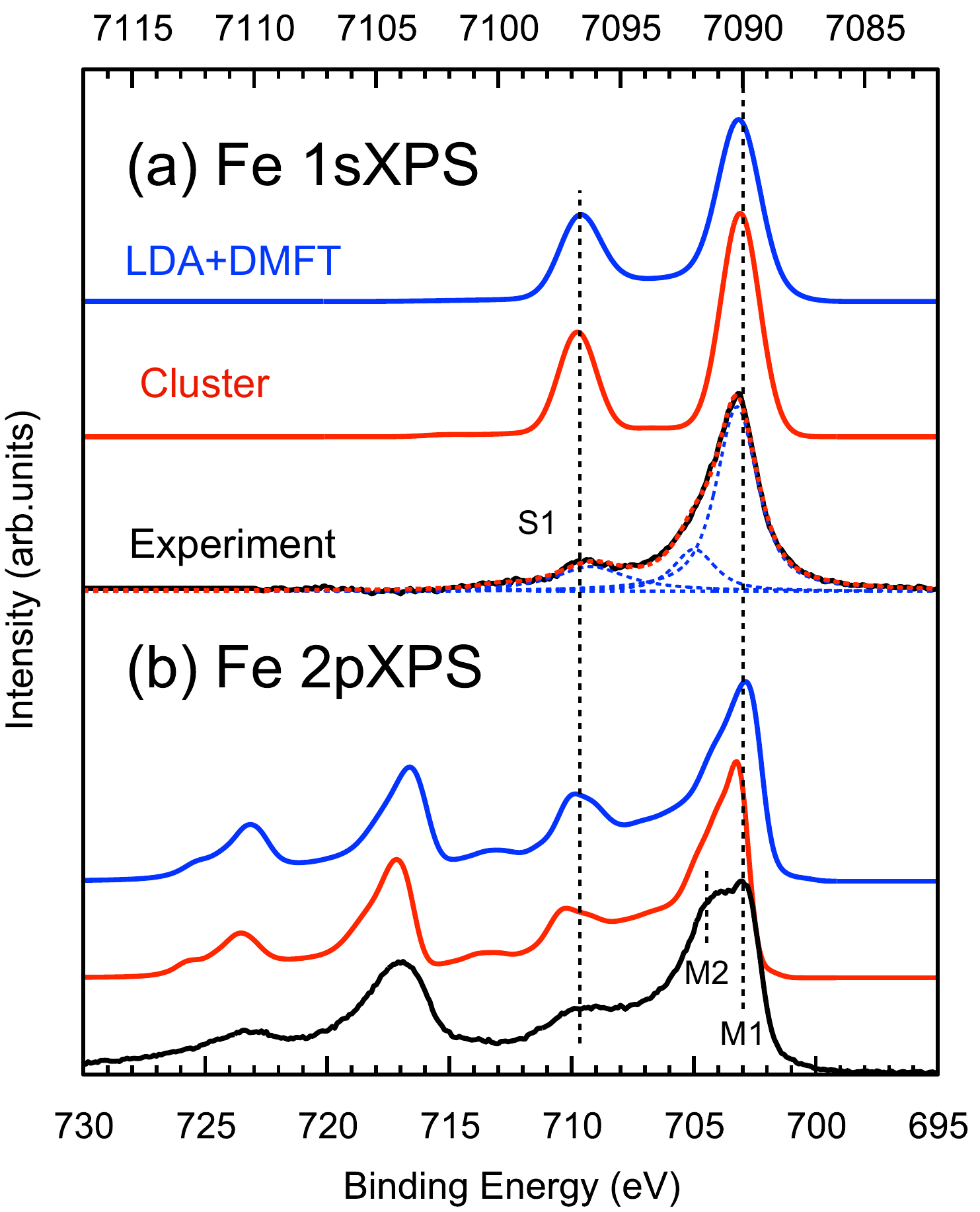}
	\caption{\textcolor{black}{(Color online) Experimental Fe 1$s$ and 2$p$  XPS spectra of FeTiO$_3$ (black) are compared with LDA+DMFT (black, solid) and cluster-model (red, solid) calculations. The spectral broadening using a Gaussian of 1.1 (0.4) eV width (HWHM) is taken into account in the calculated 1$s$ (2$p$) spectra.
	The fitting result (red, dashed) using Voigt functions (black, dashed) for the 1$s$ data is shown together.}
	}\label{fig_fto}
\end{figure}

Figure~\ref{fig_fto} shows the experimental Fe 1$s$ and 2$p$  XPS spectra. 
In the Fe 1$s$ spectrum, we observe a main line ($\sim$~7090~eV) and a CT satellite (S1~:~$\sim$~7097~eV).
The energy splitting between the CT satellite (S1) and main line is 7~eV which is about 2.5~eV smaller compared to that in Fe$_2$O$_3$, see Fig.~\ref{fig_fe}a.
As found in Table.~\ref{tab3}, the hopping amplitude between Fe and nearest-neighboring oxygen as well as other parameters does not differ so much in the two compounds.
The large difference in the main-line -- CT-satellite splitting comes from the value of the $V_{\rm eff}$, 4.19~eV for Fe$_2$O$_3$  and 3.49~eV for FeTiO$_3$ .
The $V_{\rm eff}$ value in the divalent Fe system ($d^6$) is smaller than that in the trivalent Fe system ($d^5$) due to an additional electron in the $T_{2g}$ orbital in the high-spin ground state, resulting in the observed smaller main-satellite splitting in FeTiO$_3$.
In FeTiO$_3$, a higher-$E_{\rm B}$ CT satellite (S2) is rather weak and not observed in the present data, which is 
due to little contribution of the $|d^8 \underline{v}^2 \rangle$ configuration to the ground state.
The position of S1 and the absence of S2 are well reproduced in the LDA+DMFT calculation. 

We expect that the nonlocal screening plays a minor role in FeTiO$_3$ compared to Fe$_2$O$_3$ since Ti ions, formally tetravalent $d^0$ configuration, cannot provide electrons to screen the x-ray excited Fe ion.
Simulation of the Fe 2$p$ XPS of Fe$_2$O$_3$, Fig.~\ref{fig_fe}b, revealed that nonlocal screening amplifies the intensity of M2 relative to M1.  
This is confirmed by comparing the experimental data of Fe$_2$O$_3$ and FeTiO$_3$. In FeTiO$_3$,~Fig.~\ref{fig_fto}b with weaker nonlocal screening, a smaller ratio of M2 to M1 intensities than in Fe$_2$O$_3$ is observed.
Indeed, the LDA+DMFT spectra of FeTiO$_3$ do not differ much from the cluster-model calculation, though the relative intensity of M1 and M2 is still noticeably modified by NLS.
\textcolor{black}{The main line of Fe 1$s$ XPS in FeTiO$_3$ is rather sharp compared to that in Fe$_2$O$_3$, see the fitting analysis in Fig~\ref{fig_fto} and~\ref{fig_fe}, indicating less nonlocal screening contribution. The intensity ratio of the low-energy peak ($I_1$) to     the high-energy peak ($I_2$) in the main line is $I_1$/$I_2$~=~3.92,~4.38 for Fe$_2$O$_3$ and FeTiO$_3$, respectively.
The smaller $I_1$/$I_2$ for Fe$_2$O$_3$ supports that the nonlocal screening is more effective in Fe$_2$O$_3$.}

\section{\label{sec5} Discussion }

\begin{figure}[t]
	\graphicspath{{Main/}}
	\includegraphics[width=1.0\columnwidth]{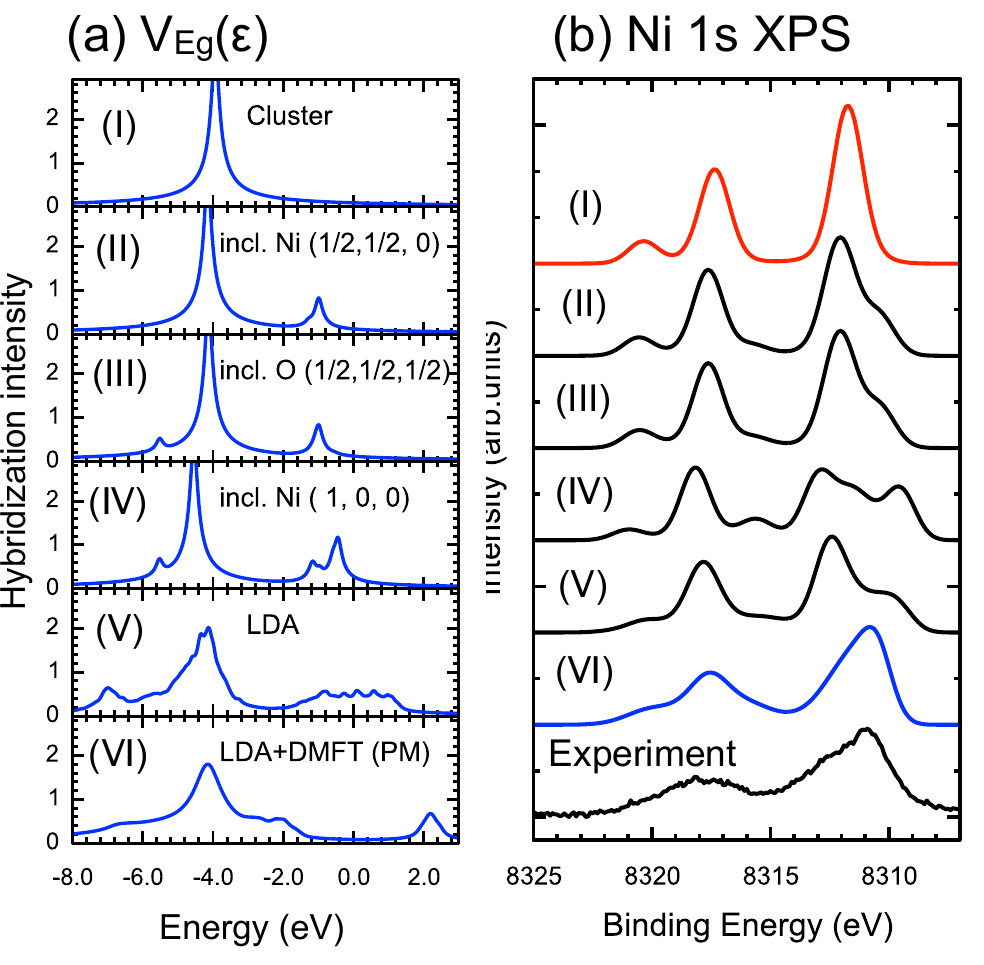}
	\caption{(Colour online)
	(a) Hybridization intensities for the $E_g$ state in NiO. From top to bottom, the long-distance hoppings including the atom denoted in the bracket are taken into account. $V(\varepsilon)$ in panels (I)-(IV) and (V) are computed in the non-interacting finite-size clusters and infinite lattice, respectively. The $V(\varepsilon)$ obtained in the LDA+DMFT calculation for the paramagnetic phase is shown in panel (VI), for comparison. The $V(\varepsilon)$ for the antiferromagnetic phase is found in Ref.~\cite{Hariki17}. 
	(b) Ni 1$s$ XPS calculated by the Anderson impurity model with the hybridization intensities in Fig.~(a).
	}\label{fig_demo}
\end{figure}

All spectral features in the studied compounds are well reproduced by the LDA+DMFT approach. This is not so for the cluster model and the comparison
of the two models provides information about the nonlocal screening effects.
Despite rather large life-time broadening of the 1$s$ spectra
compared with the 2$p$ counterparts, the charge-transfer satellites are clearly visible for the studied compounds.
This holds also for charge-transfer satellites at higher binding energies, which 
are not obscured by the overlap of spin-orbit split edges as in the 
$2p$ spectra.
The $1s$ charge-transfer satellites, usually well pronounced in the spectra of correlated insulators,
thus provide information about covalent bonding in these compounds.
Absence of the core-valence multiplets in 1$s$ XPS directly reveals the effect
of nonlocal screening reflected in the asymmetry of the 1$s$ main line.

The shape of the 1$s$ XPS spectra has implications for the interpretation of 1$s$ ($K$-edge) x-ray absorption spectroscopy (XAS). In $K$-edge XAS, the electron excited from the 1$s$ core-level to the broad 4$p$ band is not bound to the excited transition-metal atom. The fact the 1$s$ XPS spectra have multiple peaks implies that one X-ray photon energy creates a series of electrons with different kinetic energies. This is in contrast to the usual way to calculate $K$-edge XAS, i.e. it is assumed that the X-ray photon creates a single electron kinetic energy. To take the spectral shape of the 1$s$ XPS spectra into account,
the $K$-edge XAS spectra must be viewed as a convolution of the empty 4$p$ density of state (as calculated from for example multiple scattering) and the 1$s$ XPS spectrum. 
In other words, the detailed understanding of the $K$-edge XAS spectral shape requires the inclusion of 
many-body response to the core-hole potential as measured with the 1$s$ XPS spectral shape, where we note that this approach is similar in concept to the charge-transfer satellite method as applied earlier~\cite{Bair80,Collart06,Tolentino92}.
If the 1$s$ XPS spectral shape can be described by a single peak, the related $K$-edge XAS can be described from the multiple scattering of a single electron energy~\cite{Wu04}, 
As shown here, charge-transfer satellites present a sizable contribution to the 1$s$ XPS of late transition-metal oxides. 
Therefore a simultaneous analysis 1$s$ XPS and 1$s$ XAS is desirable for the detailed understanding of the 1$s$ XAS spectral shape.

Finally, we discuss the theoretical description of nonlocal screening core-level XPS. 
In contrast to the real-space approach of the multi-site cluster model~\cite{Veenendaal93}
LDA+DMFT includes both local-screening and nonlocal-screening effects in the hybridization function $V(\varepsilon)$ of the Anderson impurity model.
To see the connection between this description and the real-space one, Fig.~\ref{fig_demo}a shows the distance dependence of the hybridization intensity $V(\varepsilon)$ in NiO.
Starting from a single peak in the $V(\varepsilon)$ of the cluster model, which corresponds to the hybridization with nearest-neighboring oxygen atoms, $V(\varepsilon)$ acquires a band character by taking more distant atoms into account. 
We note that the truncated $V(\varepsilon)$ in panels (I)-(IV) and (V) is computed in non-interacting finite-size clusters and infinite lattice, respectively. 
The intensities around $-2 \sim 2$~eV corresponds to the hybridization with Ni 3$d$ bands.
These intensities are rather weak compared to those arising from O 2$p$ bands ($\sim$ $-4$~eV) due to a smaller amplitude of direct metal-metal hopping as well as indirect hopping, e.g., via a metal-ligand-metal path. The electronic correlation represented by DMFT self-energy modifies the $V(\varepsilon)$ dramatically and a gap opens at the Fermi energy. Figure~\ref{fig_demo}b shows the calculated 1$s$ XPS spectra by the Anderson impurity model with the truncated hybridization intensities $V(\varepsilon)$.
By taking surrounding Ni ions into account, a new peak develops in the low-binding-energy side of the main line.
\textcolor{black}
{This accompanies a noticeable shift of the local-screening peak ($\sim 8312$~eV) because of the following reason. In the cluster model, the main line is composed of mainly $cd^9\underline{L}$ configuration. By including the charge transfer from surrounding Ni ions, $cd^9\underline{D}$ configuration contributes the main line (here, $\underline{D}$ denotes a hole on the neighbor Ni ion). In the impurity picture, though there is no coupling between $cd^9\underline{L}$ and $cd^9\underline{D}$ configurations, indirect coupling via (unscreened) $cd^8$ configuration (i.e.~$cd^9\underline{L} \leftrightarrow cd^8 \leftrightarrow cd^9\underline{D}$) gives rise to effective repulsion between the two screened configurations.}
However we find a qualitative difference in the main-line shape between the experimental data and the spectra in (I)-(V).
The LDA+DMFT result in the paramagnetic phase, see (VI), shows double peaks in the main line although their splitting of is  narrow. The LDA+DMFT result in the antiferromagnetic  phase, see Fig.~\ref{fig_ni}b, reproduces the double-peak feature qualitatively.
The $V(\varepsilon)$ in LDA, in principle, includes hybridization with all valence states in the non-interacting lattice, indicating the importance to include the correlated Ni 3$d$ band and the magnetic ordering properly to describe the XPS spectra.

\section{\label{sec6} Conclusions }
We have studied both experimentally and theoretically the 1$s$ and 2$p$ hard x-ray photoemission spectra (XPS) in a series of late transition metal oxides:~Fe$_2$O$_3$, FeTiO$_3$, CoO and NiO.
Despite the large core-hole life-time broadening, the 1$s$ XPS benefits from the absence of core-valence multiplets and spin-orbit coupling effects in the spectra, which allows observation of high-energy satellites as well as the main-line asymmetry.  
These $1s$ XPS features can be interpreted in terms of material specific metal-ligand covalency
(satellites) and nonlocal screening (main-line asymmetry).
The 1$p$ XPS is thus complementary to 2$p$ XPS that has more complex spectra.
Using LDA+DMFT approach we were able to reproduce the $1s$ and $2p$ XPS spectra
of the studied materials, while the deviations from the cluster model allowed us to quantify
the role of nonlocal screening.
Based on the present 1$s$ XPS results, we have pointed out the importance of the 1$s$ XPS to interpret the 1$s$ ($K$-edge) x-ray absorption spectra.

\begin{acknowledgments}
The authors thank T. Uozumi, P. S. Miedema, A. Sotnikov and J. Fern\'andez Afonso for fruitful discussions, and Ties Haarman for providing the fitting code. We thank R.-P. Wang for supporting the fitting analysis.
Dimanond Light Source, Canadian Light Source, ESRF and BESSY II are acknowledged for the allocation of synchrotron radiation beamtimes at the I09, SXRMB, ID16, and KMC-1 beamlines, respectively. A. H., M.W. and J. K. are supported by the European Research Council (ERC) under the European Union's Horizon 2020 research and innovation programme (grant agreement No.~646807-EXMAG). M. G. and F. M. F. de G. are supported by an European Research Council (ERC) grant XRAyonACTIVE (No. 340279).
AR acknowledges the support from Imperial College London for her Imperial College Research Fellowship. 
The computational calculations were performed at the Vienna Scientific Cluster (VSC).
\end{acknowledgments}

\bibliography{1sXPS}

\end{document}